\documentclass[prl,twocolumn,english,showpacs]{revtex4-1}
\usepackage[utf8]{inputenc}
\usepackage[english]{babel}
\usepackage{amsmath}
\usepackage{amsfonts}
\usepackage{amssymb}
\usepackage{lipsum}
\usepackage{graphicx}

\begin{document}
	
	\title{Exponentially improved detection and correction of errors in experimental systems using neural networks}
	
	\author{Pascal Kobel, Martin Link and Michael K\"ohl}
	\affiliation{Physikalisches Institut, University of Bonn, Wegelerstra{\ss}e 8, 53115 Bonn, Germany}
	
	\begin{abstract}
		{We introduce the use of two machine learning algorithms to create an empirical model of an experimental apparatus, which is able to reduce the number of measurements necessary for generic  optimisation tasks exponentially as compared to unbiased systematic optimisation. Principal Component Analysis (PCA) can be used to reduce the degrees of freedom in cases for which a rudimentary model describing the data exists. We further demonstrate the use of an Artificial Neural Network (ANN) for tasks where a model is not known. This makes the presented method applicable to a broad range of different optimisation tasks covering multiple fields of experimental physics. We demonstrate both algorithms at the example of detecting and compensating stray electric fields in an ion trap and achieve a successful compensation  with an exponentially reduced  amount of data.} 
	\end{abstract}
	
	\maketitle

The successful acquisition of data from experiments requires a careful calibration and correction of environmental influences. Across the fields of atomic and condensed matter physics, the continuously  increasing requirements regarding the precision of experiments, for example, in atomic clocks and quantum computers requires an ever better cancellation of disturbances in order to maintain quantum coherence. Similarly, advanced experiments  in particle physics  rely on more accurate stabilization of the position and focus of particle beams.  Hence, detecting and reacting to an external noise source, which drives an experimental system out of its optimal state, and compensating it, is a frequent challenge for many experimental platforms. This can be particularly cumbersome when the timescale required for readjustment of the setup is comparable to the timescale of the temporal deviation of the noise source. Therefore, efficient and fast methods for the detection and compensation of external perturbations are of high importance. Even though the physical sources of perturbations can be quite different and specific to an experimental setup, for example, time-varying stray electromagnetic fields over thermal drifts to acceleration and motion of the apparatus,  the optimization problem is fairly generic.

Of course, a comprehensive simulation of the whole experimental system including all possible fringe effects would enable the prediction of deviations from the optimal state due to external deviations. However, such a simulation is usually impossible since it would require deep knowledge of all special characteristics of the setup and  the noise sources. When weighing up effort and gain of a universally applicable simulation compared to a unbiased systematic search for an optimal setting within a (sometimes huge) parameter space, the latter wins in most cases. A potentially smart way out could originate from  machine learning research, which in recent years has found its way into the domain of physics \cite{RevModPhys.91.045002}. For example, reinforcement learning techniques  have been  utilized to systematically scan large potential parameter spaces in search for optimal values \cite{Barker_2020, Wigley:2016aa, Tranter:2018aa,wu2020active} and supervised learning techniques yielded improved measurements and optimised critical parameters \cite{Mavadia:2017aa, PhysRevA.95.012335, Schuff_2020}.

\begin{figure}
\centering
\includegraphics[width=\columnwidth]{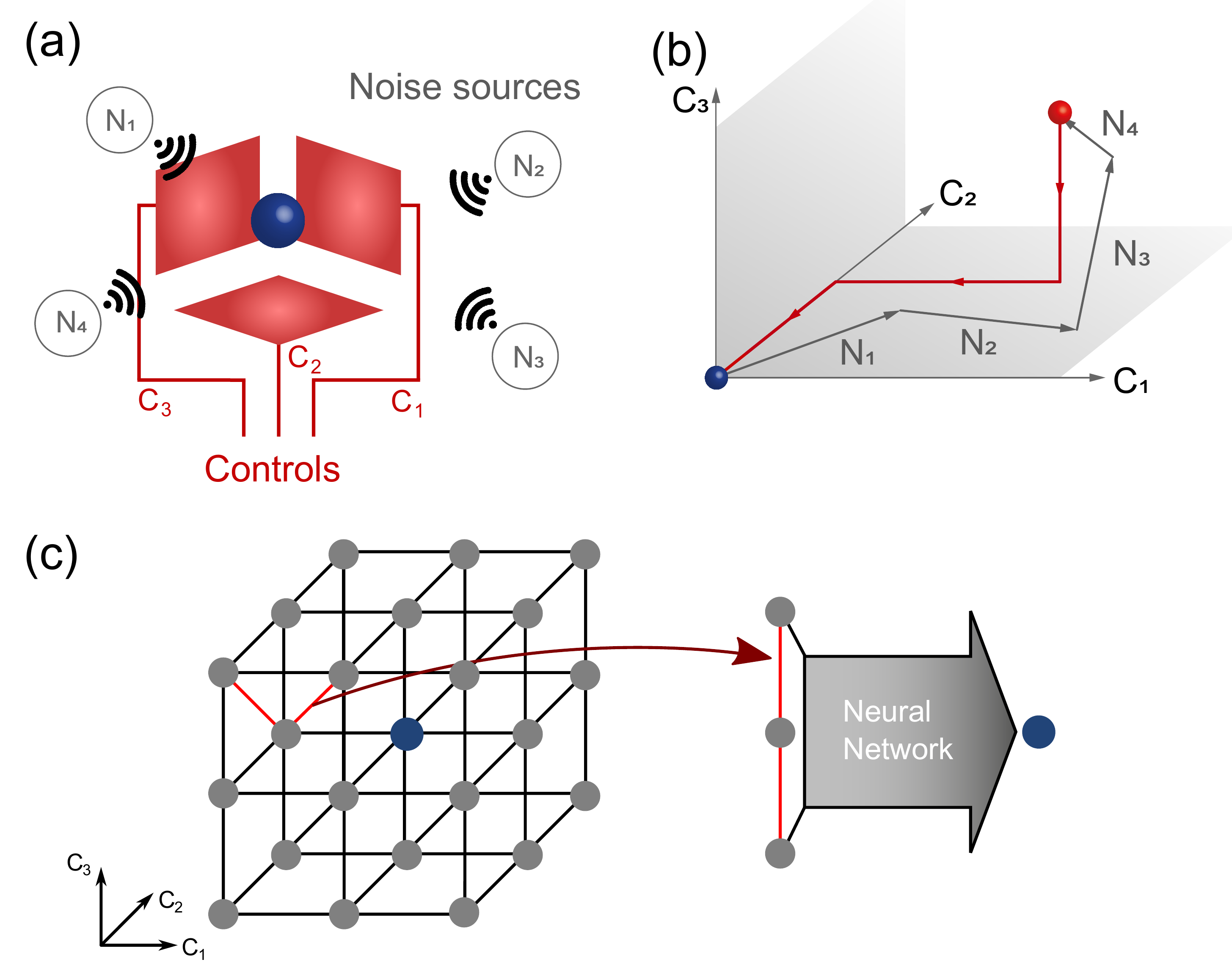}
\caption{\textbf{(a)} An experimental system (blue circle) is disturbed by external noise sources $\{N_1,... N_4\}$. Experimental controls  $\{C_1,...C_3\}$ along counteract these disturbances with $N$ parameters. In general, the compensation axes do not coincide with the axes of disturbance. \textbf{(b)} The noise sources  translate the operating point away from the optimal  point on a trajectory in $k$-dimensional parameter space. The controls should cancel the noise sources. \textbf{(c)} Concept of the optimization: The volume of the hypercube  $n^k$ is sampled along lines across the hypercube, which are fed into a neural network.  } \end{figure}

Here, we demonstrate a different machine learning approach to greatly simplify the correction of error sources without the need of a lengthy exploration through parameter space. We experimentally demonstrate our novel methodology at the example of the electric stray-field compensation of a radiofrequency ion trap. By applying supervised and unsupervised machine learning algorithms, we are able to create an accurate empirical model of the underlying experimental system, which allows us to access the correlations between parameters, for example, induced by the geometry of the setup.  The machine learning approach has the advantage to work directly on existing data and does not need any additional information to compensate an external perturbation with an accuracy which is limited only by measurement error. We demonstrate that our approach reduces the number of required measurements to reach the same accuracy as compared to an unbiased systematic search  by several orders of magnitude and features an exceptionally better scaling behaviour.



The concept of our approach is shown in Figure 1a. We consider an experimental system (in blue), which is perturbed by a set of noise sources $\{N_1,N_2,...N_m\}$ in unspecified locations, i.e., with an unknown cross-correlation between them. The noise sources can be static or time-dependent and they drive the experimental system out of its optimal operating condition. This corresponds to a displacement of the system in a parameter space, see Figure 1b. The goal is the cancellation of the noise sources at the location of the experiment with a set of linearly independent controls $\{C_1, C_2, ..., C_k\}$. The  value of $k$ denotes the dimensionality of the configuration space in which the optimization takes place. The $ k $--dimensional set of control variables which brings the system back to the optimal operation conditions cancels the external noise sources and is referred to as compensation point. 
Without any prior knowledge, the optimisation routine would need to scan the whole configuration space, say, on a hypercubic grid with $n$ datapoints along $k$ directions, i.e., order $n^k$ datapoints, see Figure 1c. The value of $n$ depends on the required precision. 

Here, we take a different approach, see Figure 1c: We feed line segments across the cubic grid into a neural network and train the neural network with the data to predict the optimal  point at which the effect of the noise sources is canceled. The input into the neural network is a set of numbers, representing several points in the $k$-dimensional parameter space. We have found the neural network to perform this optimization task very efficiently and hence, we choose a linear benchmark for the number of measurements  $a \cdot n \cdot k$ required for a successful prediction of the compensation point. Here, $n \cdot k$ represents the number of points of one line segment in every dimension within the hypercube and $a$ measures the improvement due to the neural network and ideally  is $a<1$. In the following, we make use of two distinct algorithms: the principal component analysis (PCA) and an artificial neural network (ANN) and show that both of them outperform a linear model. We furthermore show that by taking accuracy measures into account, the amount of input data can be reduced to a minimum in the case of the ANN, which is even below the principal minimum of $k$ data points of the PCA.

The Principle Component Analysis (PCA) \cite{pearson1901, hotelling36} is an unsupervised learning algorithm which can identify the direction of maximum variance within a dataset in parameter space. In the best case, all deviations from the optimally compensated experiment can be explained by one principal component.   Then, the multiplicative factor $a_\text{PCA}$ approaches $a_\text{PCA} = 1/k$. This can be expected for an electric or magnetic field perturbing a very small object such that field gradients can be neglected.  Using the PCA it is possible to determine this field axis and the task of experimental optimisation can be reduced to a single component in a transformed parameter space. However, the PCA method works also for more general and more complicated deviations as long as we can provide a linear model of the problem. 

To go even further, we use a fully-connected artificial neural network (ANN) to train on compensation data using back-propagation \cite{rumel86, rumel86b} and Adam gradient descent optimization \cite{adamGD}. In comparison to PCA, this approach has no need for a suited data representation and can capture non-linear transformations as well. By training the ANN with just the equivalent of a few full compensation runs of data, it is possible to push the network prediction  accuracy to the limit of the experimental accuracy of the training dataset. The coefficient  $a_\text{ANN}$ depends on the desired accuracy. We find empirically that the neural network exceeds the performance of the PCA with  $1/(n k) \leq a_\text{ANN} \leq a_\text{PCA}$ and note that the ANN can perform particularly efficient when a lower prediction accuracy than the underlying training dataset accuracy is needed. 


\begin{figure}
\centering

\includegraphics[width=\columnwidth]{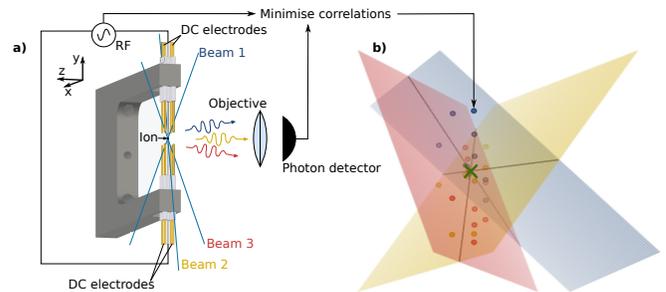}
\caption{Schematic of ion trap setup and compensation of stray fields. (a) Three non-collinear laser beams illuminate the trapped ion and the fluorescence counts are detected on a single-photon counter. The correlation between photon counts and supplied radiofrequency  measures the strength of the electric stray field.  (b) Upon varying the control voltages supplied to the dc electrodes, all points of minimal detected correlations lie on a plane for each beam. The intersection of the planes corresponding to the different beams is the optimal stray fields compensation (denoted by X). 
}
\label{figureTrapSetup}
\end{figure}

	In the following, we demonstrate an application of the concept presented in the previous section and demonstrate its superiority over previous methodology. We target the compensation of electric stray fields in a Paul trap confining a single trapped ion. The Paul trap comprises of an electric quadrupole field oscillating at frequency $\Omega$ which  confines the ion at its center, see Figure 2a. However, stray electric fields push the ion away from the symmetry point of the quadrupole field and increase the (micro--)motion of the ion, which severely limit its use in quantum information processing \cite{Blatt2008} and atomic clocks \cite{doi:10.1063/1.4930037,PhysRevLett.104.070802}. We compensate stray fields along an arbitrary direction by using a linear combination of electric fields applied in  three orthogonal directions. Hence, the  dimensionality of the compensation parameter space is $k=3$.

There are several established methods for detecting  stray electric fields by probing  the micromotion of a trapped ion \cite{PhysRevLett.94.230801,FirstDopplerCorrelationThreeBeamExample,VariousRFPowerIonTrajectory,DetectionOfParametricResonance,doi:10.1063/1.3665647,parametricExcitation}. Here, we utilize the Doppler correlation technique \cite{segmentedPaultrap,PhysRevA.52.2994,FirstDopplerCorrelationThreeBeamExample}. To this end, we illuminate the ion with a laser beam and record the scattered photons on a single-photon counter, see Figure \ref{figureTrapSetup}. The correlation between photon arrival times and supplied radiofrequency $ \Omega $ measure the excess motion of the ion along the laser beam axis. This methodology, however, provides only information about the motion of the ion with a component along the wave vector of the laser beam, whereas motion of the ion in the orthogonal directions cannot be detected. Hence, a full three-dimensional sampling of the electric field compensation requires three non-colinear laser beams. Unfortunately, in practice the minimisation of motion along one beam axis likely introduces additional motion along the two remaining beam axes. This non--trivial coupling is determined by the relative orientation of the laser beams, trap geometry and direction of applied electric dc fields. 
One can map the configuration space in a cubic grid as in Figure 1c by performing $n^k$ measurements of the ions's motion resulting a systematic search in the parameter space of applied DC electric field \cite{PhysRevA.87.013437}. 
Iterative methods \cite{FirstDopplerCorrelationThreeBeamExample} or data describing mathematical models like trajectory analysis \cite{VariousRFPowerIonTrajectory} can reduce the number of required measurements for a full compensation, however, no previously reported method has reached a better scaling than $ a\cdot n\cdot k $ with $ a\cdot n\geq 3 $ because they  capture only parts of the coupling between the parameters and nature of the disturbance.

In the Doppler correlation measurement, all points for which we detect minimal motion along one probe laser beam axis, form a hyperplane in parameter space for the investigated region. The orientation of the plane depends on the coupling between the experimental parameters as described above and needs not to be orthogonal to the corresponding beam axis. The parameter set for which all three planes intersect is the ideal compensation point. In order to test the accuracy of the prediction of our optimization procedure, we define a measure $\sigma$ as the standard deviation of predicted compensation points for several input data sets of a common noise manifestation in terms of their Euclidean distance in parameter space. We have verified that the predicted compensation is in 68$\%$  of the cases within one standard deviation from the actual one. From a pure geometric point of view, determining the parameters defining a plane requires a minimum of three correlation measurements per plane. The minimal approach of nine measurements provides us with $\sigma=734$\,V/m and therefore the accuracy is worse than the typical range of compensation drift, which averages to $\sim600$\,V/m over a few months. The reason for this short-fall is that the quality of the fitted plane parameters and hence the determination of their crossing depends critically on the number of datapoints per plane. In practice, we have found that the we require a minimum of $n=8$ points per plane (i.e. 24 in total) to calculate the point of intersection with an accuracy of 53\,V/m, which is sufficient for our purposes (see Figure 3). 

Next, we extend the plane--model by a principal component analysis (PCA) with the goal of further reducing the number of datapoints needed for finding the intersection point. We achieve a full compensation with 1 data point  per beam (i.e. $  a_\text{PCA}\cdot n = 1$), which, remarkably, is below the minimal requirement of measurements to mathematically define a plane. With the self-learned correlations, which previously were not covered by our data model, we are able to predict the correct compensation point with an accuracy of $\sigma=42$\,V/m. The PCA-enhanced plane model reduces the number of measurements required by one order of magnitude, from 24 to 3, with even improved values of $\sigma$. This advantage originates in the PCA method's capability of revealing possible correlations between the plane parameters without knowing the final optimal compensation point. Therefore, a suited linear combination of the plane parameters represented by the first principal component (which represent 89\% of variance) leads to a reduction of required measurements. Moreover, we find that $\sigma_\text{PCA}$  is even slightly smaller than our measurement accuracy of 54\,V/m. This shows that the limiting factor for prediction accuracy is not the method of machine learning, but rather the measurement accuracy within the experiment. This, however, is quite a generic situation as any model of noise sources is always approximate and machine learning algorithms can efficiently spot correlations which have been unknown or are incorrectly captured by the model. 

\begin{figure*}
	\includegraphics[width=0.6\textwidth]{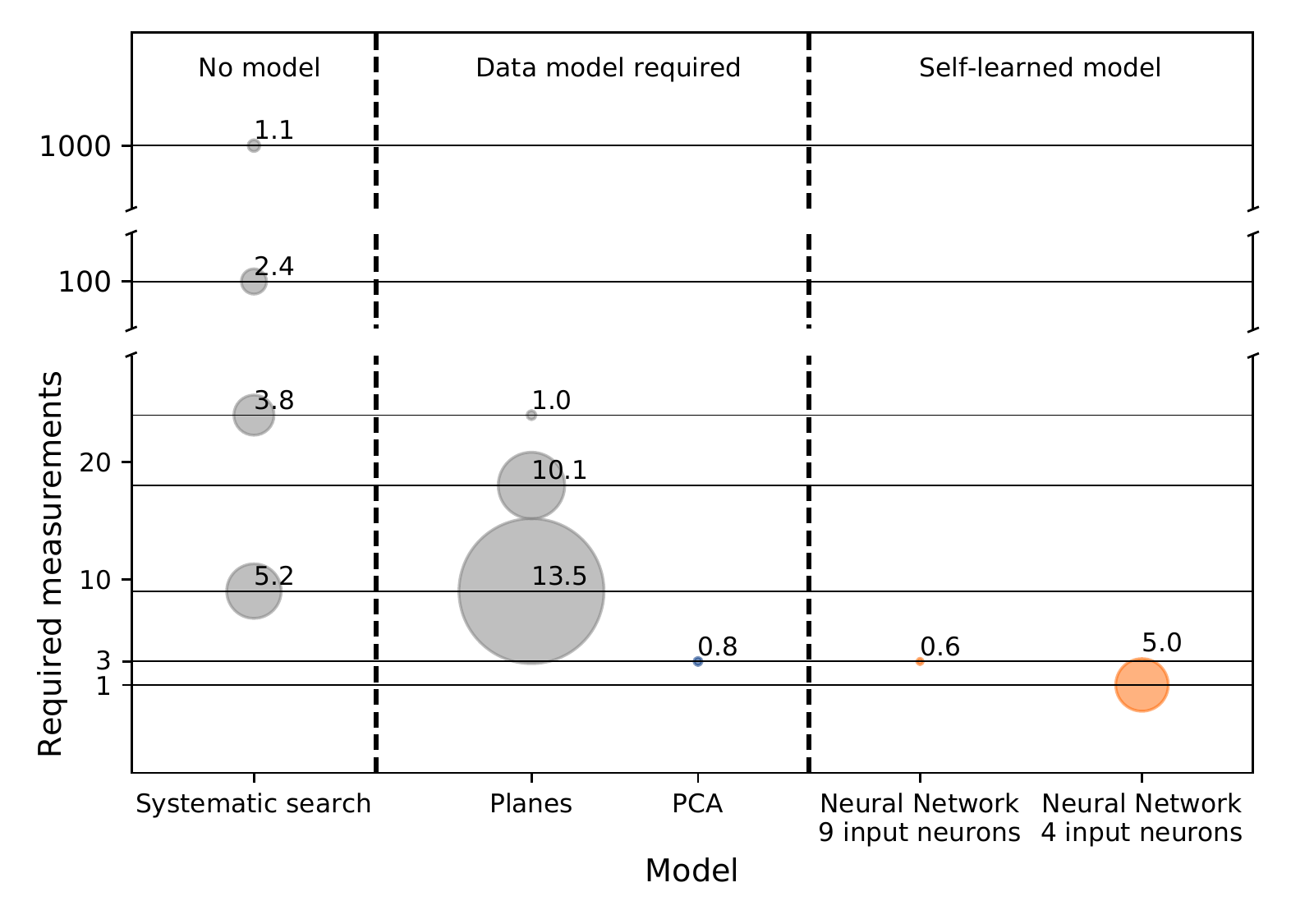}
	\caption{Comparison of the accuracy of the different compensation methods. The size of the circles denotes the standard deviation of the spread $ \sigma $ (diameter of the circles) normalised to the measurement accuracy of $54$\,V/m. For reference, over three months of operation the spread of measured compensation points amounts to 600\,V/m which reflects the changing environmental conditions in the apparatus. 
}
	\label{disturbance}
\end{figure*}

In order to even eliminate the need for the pre-existence of any mathematical model, we choose a neural network to exploit possible hidden correlations to decrease the amount of data needed for prediction beyond linear sets of parameters. The fully-connected neural network consist of a hidden layer with a size of 16 neurons and 3 output neurons representing the 3 spatial components of the compensation point. For the input layer we compare two different sizes, namely 9 neurons representing three points in parameter space (one point per beam) and 4 neurons, representing one point in parameter space and the identification of the beam $ i $ along which the ions's motion was minimised $ i\in \{1, 2 , 3\}.$ In contrast to the self--learning PCA algorithm, for training of the neural network, we have to provide "correct" data to learn from. We use data from seven full compensation measurements, each representing a different charge distribution in our setup and consisting in total out of 137 datapoints. In Figure \ref{disturbance} we show that already this comparably small amount of data is sufficient to train the network, such that the compensation point can be predicted on an unknown dataset with  $ \sigma_{\text{ANN},{9}} =34 $\,V/m for an input layer of 9 neurons. This compares to the PCA--enhanced plane model in accuracy and amount of needed data. However, the big advantage of this method is that it does not take any model such as the plane--model into account and is therefore applicable even in cases where no model for the noise sources and coupling between experimental parameters exists or is known. In the case of the 4-neuron-neural network, the accuracy drops by about $30\,\%$ and shows that the provided information of one datapoint is not sufficient to create a model as precise as for a 9 neuron-input. However, in cases where lower accuracy is acceptable constitutes an additional factor of 3 reduction in needed input data.

In Figure 3 we show the results of the prediction accuracy of the discussed models. The unbiased systematic sampling shows the slowest convergence, as a function of the number of measurements, reaching a compensation uncertainty of 60\,V/m only after 1,000 measurements. For comparison, the model using the intersecting planes reaches a comparable accuracy already after 24 measurements. Even though this gain is quite impressive, the machine learning algorithms achieved a further reduction in measurements of  one order of magnitude. Furthermore, the reduction to only one measurement along one beam-axis in case of the 4-neuron model leads to a significant simplification of the measurement process. Further, in this case we achieve a remarkable reduction of $ a_{\text{ANN},4}\cdot n \cdot k =1  $ which means, that our system requires a smaller number of datapoints than the dimension $ k $ of the problem. This is comparable to a perfect $ k $-dimensional mathematical model where just one unique point is enough to complete the whole model and derive the compensation point from, except for the fact that with the ANN we only have access to this single compensation point instead of the whole point-space. The use of two hidden layers (each 16 neurons) for the neural networks brought no further improvement.


In conclusion, we have shown that machine learning techniques can greatly help to perform generic optimisation tasks leading to an exponential speed-up of calibration measurements.  Not only does this speed up data acquisition, but might also facilitate active stabilisation of noise sources which previously was impossible due to too long data acquisition and analysis. Although, generally, machine learning approaches require large amounts of training data, we show that even very little data can achieve a similar or better prediction accuracy compared to a pre-defined data describing mathematical model solely limited by the quality of data provided to the machine learning algorithms. We have demonstrated our approach on a generic optimisation problem  and expect that our  methodology is widely applicable to a broad range of experimental settings.   

P.K. and M.L. contributed equally to this work. We thank M. Breyer, K. Kluge, H.-M. Meyer, and V. Sair for their contributions and discussions. This work has been supported by BCGS, the Alexander-von-Humboldt Stiftung, DFG (SFB/TR 185 project A2), by the Deutsche Forschungsgemeinschaft (DFG, German Research Foundation) under Germany's Excellence Strategy – Cluster of Excellence Matter and Light for Quantum Computing (ML4Q) EXC 2004/1 – 390534769, and  BMBF (FaResQ and Q.Link.X).

\end{document}